\documentclass[preprintnumbers,amsmath,amssymb,nofootinbib]{revtex4}
\usepackage{graphicx}
\usepackage{dcolumn}
\usepackage{bm}
\usepackage[dvips]{color}

\newcommand{\beq}{\begin{equation}}
\newcommand{\eeq}{\end{equation}}
\newcommand{\bq}{\begin{equation}}
\newcommand{\eq}{\end{equation}}
\newcommand{\ba}{\begin{array}}
\newcommand{\ea}{\end{array}}
\newcommand{\beqa}{\begin{eqnarray}}
\newcommand{\eeqa}{\end{eqnarray}}

\def\[{\left[}
\def\]{\right]}
\def\({\left(}
\def\){\right)}

\evensidemargin=\oddsidemargin

\def\pslash{\not{\hbox{\kern-4pt $p$}}}
\def\qslash{\not{\hbox{\kern-4pt $q$}}}
\def\lv{\not{\hbox{\kern-4pt $L$}}}
\def\lsim{\mathrel{\raise.3ex\hbox{$<$\kern-.75em\lower1ex\hbox{$\sim$}}}}
\def\gsim{\mathrel{\raise.3ex\hbox{$>$\kern-.75em\lower1ex\hbox{$\sim$}}}}
\def\ifmath#1{\relax\ifmmode #1\else $#1$\fi}
\begin{document}

 \title{Low-Energy Effective Theory, Unitarity, and Non-Decoupling Behavior\\
 in a Model with Heavy  Higgs-Triplet Fields}

\author{ R.\ Sekhar Chivukula, Neil D. Christensen,  and Elizabeth H.\ Simmons}

 \affiliation{
 Department of Physics and Astronomy, Michigan State
       University, East Lansing, MI 48824, USA}

\date{\today}

\begin{abstract}
We discuss the properties of a model incorporating
both a scalar electroweak Higgs doublet and an electroweak Higgs triplet. 
We construct the low-energy effective theory
for the light Higgs-doublet in the limit of small (but nonzero) deviations in
the  $\rho$ parameter from one, a limit in which the triplet states become heavy.
For $\Delta \rho >  0$, perturbative unitarity of $WW$
scattering breaks down at a scale inversely proportional to the renormalized
vacuum expectation value of the triplet field (or, equivalently, inversely proportional
to the square-root of $\Delta \rho$).  This result  
imposes an upper limit on the mass-scale of the heavy triplet bosons
in a perturbative theory; we show that this upper bound is consistent
with dimensional analysis in the low-energy effective theory.
Recent articles have shown that the triplet bosons do not decouple, in
the sense that deviations in the $\rho$ parameter from one do not
necessarily  vanish at one-loop in the limit of large triplet mass.
We clarify that, despite the non-decoupling behavior of the Higgs-triplet, 
this model does not violate the decoupling theorem since it
incorporates a large dimensionful coupling.
Nonetheless, we show that if the triplet-Higgs boson masses are of order the GUT scale, 
perturbative consistency of the theory requires 
the (properly renormalized) Higgs-triplet vacuum expectation
value to be so small as to be irrelevant for electroweak phenomenology.

\end{abstract}


\date{December 4, 2007}
 
 \maketitle

\section{Introduction}

Recent articles on a model incorporating both a scalar electroweak Higgs doublet and an electroweak Higgs triplet \cite{Passarino:1989py,Passarino:1990nu,Lynn:1990zk,Blank:1997qa} have established that the model exhibits non-decoupling behavior \cite{Chen:2005jx,Chen:2006pb,Chankowski:2006hs,Chankowski:2007mf}, in the sense that deviations in the $\rho$ parameter from one do not
necessarily  vanish at one-loop in the limit of large triplet mass.  Such behavior appears  difficult to reconcile with the expectation that the effects of the heavy fields should be suppressed
by inverse powers of their mass. This presents a puzzle since the only renormalizable theory with a scalar Higgs doublet has a $\rho$ parameter of precisely one (at tree-level). 

In this paper, we explicitly construct the low-energy effective field theory of this model obtained by integrating out the Higgs triplet -- thereby showing that there exists a perfectly sensible low-energy effective theory with a consistent dimensional analysis scheme. 
We find that the higher-dimensional, non-renormalizable, operators responsible
for deviations in $\rho$ are suppressed not by inverse powers of the triplet mass, but rather
by powers of the renormalized triplet vacuum expectation value (vev) divided by the renormalized
doublet vacuum expectation value. Therefore, the low-energy theory is perturbative only up to a scale which is inversely proportional to the renormalized triplet vev (or, equivalently, inversely proportional
to the square-root of $\Delta \rho$). 
We show that two possibilities remain: either the contribution of the triplet vev is comparable to the existing experimental bounds, in which case the triplet scalars must have a mass of order 30 TeV or
lower, or if the triplet masses are very heavy (much larger than 30 TeV) then the triplet vev is too small to be phenomenologically relevant. 

Some authors \cite{Chankowski:2006hs,Chankowski:2007mf} have postulated that the model violates the Appelquist-Carazzone decoupling theorem \cite{Appelquist:1974tg}.  We clarify that the presence of a large dimensionful coupling in this model implies that the decoupling theorem simply is not applicable in this case.

After introducing the full model in Section \ref{sec:two}, we determine the mass-eigenstate fields and show that in the limit of small weak-isospin violation ($\Delta \rho \ll 1$) the spectrum consists approximately of a light Higgs-doublet and a heavy Higgs-triplet field.   We then integrate out the heavy states to obtain a tree-level low-energy effective theory of the light states, which are dominantly composed of the original Higgs-doublet states. We discuss the extension to higher loop order and use
dimensional analysis to argue that the value of  $\Delta\rho$ places an upper bound on the mass of the heavy mostly-triplet states -- essentially because inclusion of these heavy states is the high-energy completion of the low-energy effective theory.  In Section III, we make that mass bound more precise by analyzing perturbative unitarity in $W^+_L W^-_L$  scattering, first at tree level and then at higher order.   Section IV discusses the non-decoupling behavior of the triplet states in the context of the effective
 field theory and shows that it arises only in the limit where a dimensionful coupling becomes large.  This makes clear that the absence of decoupling is not a violation of the decoupling theorem \cite{Appelquist:1974tg}.

We then turn to the question of whether non-decoupling implies there must be low-energy consequences of the presence of the heavy fields. We find that, in order for the low- and high-energy theories to both be perturbative, one must adjust the renormalized value of the triplet
vev to be of order $v^2/\Lambda$, where $v=(\sqrt{2}G_F)^{-1/2}\approx 250$ GeV and $\Lambda$ is the mass scale associated with the high-energy completion.  If $\Lambda$
is greater than about 30 TeV, this is more stringent than tuning the triplet vev to produce a phenomenologically acceptable Higgs-triplet contribution to $\Delta\rho$.  As a result, in the context of a theory with a high completion scale, like a non-supersymmetric grand-unified theory (GUT) \cite{Georgi:1974sy}, the constraint on the triplet vev is so severe that heavy Higgs-triplet fields would not produce any experimentally-visible consequences -- despite their intrinsic ``non-decoupling" nature.

\section{The Low-Energy Effective Theory}
\label{sec:two}

\subsection{The Model}
\label{sec:model}

We will focus on the gauge- and scalar-sectors of an $SU(2)_W \times U(1)$ model
\cite{Passarino:1989py,Passarino:1990nu,Lynn:1990zk,Blank:1997qa}
with a complex scalar doublet $H$, which transforms as a $2_{+\frac{1}{2}}$,
and a real triplet field $T^a$ which transforms as a $3_0$. For the triplet
field we will use the $2 \times 2$ matrix
\begin{equation}
T = \frac{T^a \sigma^a}{2}~,
\end{equation}
where the $\sigma^a$ are the usual Pauli matrices. Under an $SU(2)_W$ transformation
$L$, these fields transform as
\begin{equation}
H \to L\, H \ \ \ \ \ T \to L\, T\, L^\dagger~.
\label{eq:gauge}
\end{equation}
The scalar part of the
Lagrangian for this model may be written
\begin{equation}
{\cal L}_{scalar} = D^\mu H^\dagger D_\mu H + 
{tr}\,D^\mu T D_\mu T - V(H, T)~,
\label{eq:lagrangian}
\end{equation}
where the most general renormalizable potential is given by
\begin{align}
V(H,T) & = m^2_H H^\dagger H + \frac{\lambda_H}{4}(H^\dagger H)^2 + 
m^2_T \,{tr} \,T^2 \nonumber \\
& + \frac{\lambda_T}{4} ({tr}\, T^2)^2 + \kappa H^\dagger H ({tr}\,T^2)
+\mu H^\dagger T H~.
\label{eq:potentiali}
\end{align}
Note the presence of the dimensionful coupling $\mu$;
we can absorb the sign of $\mu$ into $T$ -- and by convention, we will take
$\mu > 0$. 
For future reference, we also note that the scalar self-couplings $\lambda_H$, $\lambda_T$,
and the $H$-$T$ coupling $\kappa$ must all be smaller than $(4\pi)^2$ in order for
the theory to remain perturbative.
The form of the potential given above is convenient for matching the model
to the expectations of, for instance, an $SU(5)$ grand unified theory \cite{Georgi:1974sy}:
here the electroweak triplet arises \cite{Senjanovic:1979yq,Chankowski:2007mf} from the 
{\bf 24} of $SU(5)$ and one expects
$\mu$ and $m_T$ to be of order\footnote{Of course, one expects $m_H$
also to be of order $M_{GUT}$ -- this is the ordinary gauge hierarchy problem 
\protect\cite{Gildener:1976ai,Gildener:1976ih}.} $M_{GUT}$.

While Eq. {\ref{eq:potentiali}} is useful in discussing the origin of the
dimensionful terms in the potential, it is not the most convenient form in which to
examine electroweak symmetry breaking. Rather,
we begin our analysis by rewriting this potential in the form
\begin{align}
V(H,T) & =  \alpha\,{tr}{T}+ \frac{\tilde{\lambda}_H}{4} \left(H^\dagger H - \frac{v^2_H}{2}\right)^2
+ \frac{\lambda_T}{4}\left({tr}\, T^2 - \frac{v^2_T}{2}\right)^2 \nonumber \\
& + \kappa \left(H^\dagger H - \frac{v^2_H}{2}\right) \left({tr}\, T^2 - \frac{v^2_T}{2}\right) \nonumber \\
& + \frac{\mu v_T}{v^2_T}\, {tr}\left[\frac{v_T}{v_H}  H H^\dagger + 
\frac{v_H}{2} T -\frac{v_Hv_T}{4}\,{\cal I}\right]^2~,
\label{eq:potentialii}
\end{align}
where ${\cal I}$ is the $2 \times 2$ identity matrix and the reason for writing the coefficient
of the  last term as $(\mu v_T / v_T^2)$ will become apparent later. Note that the  last term is gauge-invariant
because of the transformation laws in Eq. \ref{eq:gauge} and,
by convention, we take $v_T$ positive. The term $\alpha\,{tr}\,T$ is a Lagrange multiplier that
will be useful in the calculations below to impose the constraint that $T$ is traceless.
Up to an irrelevant constant and the addition of the Lagrange multiplier, 
Eqs. \ref{eq:potentiali} and \ref{eq:potentialii} are
the same with the identification
\begin{align}
m^2_H & = -\,\frac{\tilde{\lambda}_H v^2_H}{4} -\, \frac{\kappa v^2_T}{2}-\frac{\mu v_T}{2}~, 
\label{eq:msquaredH}\\
m^2_T & = -\, \frac{\lambda_T v^2_T}{4} -\, \frac{\kappa v^2_H}{2} + \frac{\mu v^2_H}{4v_T}~,
\label{eq:msquaredT}\\
\lambda_H & = \tilde{\lambda}_H +\frac{4 \mu v_T}{v^2_H}~.
\label{eq:lambdah}
\end{align}
Note the appearance of the term $4\mu v_T/v^2_H$ in the relationship between
$\lambda_H$ and $\tilde{\lambda}_H$ -- this is the first manifestation of how a {\it ratio} involving a
dimensionful coupling ($\mu$) can appear in what would otherwise look like a simple dimensionless
coupling.  We will show that this behavior prevents the model from satisfying 
the conditions necessary for the applicability 
of the Appelquist-Carazzone decoupling theorem \cite{Appelquist:1974tg}; the 
non-decoupling behavior of this model 
\cite{Chen:2005jx,Chen:2006pb,Chankowski:2006hs,Chankowski:2007mf} is discussed 
in section \ref{sec:decoupling}.

As written in Eq. \ref{eq:potentialii} the potential is positive semi-definite so
long as $\tilde{\lambda}_H$ and $\lambda_T$ are positive, and
\begin{equation}
\tilde{\lambda}_H \lambda_T \ge 4 \kappa^2~.
\end{equation}
The global minimum of the potential, therefore, corresponds to the vacuum
expectation values (vevs)
\begin{equation}
\langle H \rangle = \begin{pmatrix}
0 \cr
\frac{v_H}{\sqrt{2}}
\end{pmatrix}
\hskip0.75in
\langle T \rangle = 
\begin{pmatrix}
\frac{v_T}{2} & 0 \cr
0 & -\,\frac{v_T}{2}
\end{pmatrix}~.
\end{equation}
These vevs yield the $W$ and  $Z$-boson masses
\begin{equation}
M^2_W = \frac{e^2}{4 s_W^2} \left(v_H^2 + 4 {v_T}^2\right)~,
\ \ \ \ 
M^2_Z = \frac{e^2\, v_H^2}{4s_W^2 c_W^2}~,
\label{eq:mwz}
\end{equation}
where the $SU(2)_W$ coupling is given by $e/s_W$, the $U(1)_Y$
coupling by $e/c_W$, $e$ is the electric charge and $s_W$ and $c_W$ are the sine and cosine
of the weak mixing angle.
Following \cite{Blank:1997qa,Chen:2005jx,Chen:2006pb}, it is convenient to define
\begin{equation}
v=\sqrt{v^2_H + 4 v^2_T}=(\sqrt{2}G_F)^{-1/2}\approx 250\, {\rm GeV}~,
\label{eq:v}
\end{equation}
and an angle $\delta$ such that 
\begin{equation}
v_H=v \cos \delta~, \ \ \ \ v_T =\frac{v}{2}\, \sin \delta~.
\label{eq:sindelta}
\end{equation}
Hence, we find the tree-level relation
\begin{equation}
\Delta \rho=\frac{M^2_W}{M^2_Z c_W^2}-1 = \frac{4 {v_T}^2}{v_H^2} = \tan^2 \delta~.
\label{eq:deltarho}
\end{equation}

\subsection{Masses and Mixing Angles}
\label{sec:massmix}

In order to determine the mass eigenstates of the theory, we first need to specify the limits in which the theory makes phenomenological sense. Experimentally, we know that $\Delta \rho \ll 1$ -- therefore, $v_H \simeq v$, and it is reasonable to
expand observables in powers of $v_T/v_H$. We will also need to decide
how to treat the dimensionful coupling $\mu$ -- we will choose to keep terms
of order $\mu v_T$ ({\it i.e.} we will assume $\mu v_T/v^2_H \le {\cal O}(1)$).
As we will see, it is inconsistent to assume $\mu$ grows any faster than $1/v_T$  in 
the small-$v_T$ limit: this constraint can be viewed as the result of ensuring that the four-point
couplings  remain perturbative in both the low- and high-energy theories (see the discussion following
Eq. \ref{eq:lambdahi}).

With these issues in mind, we may proceed to determine the mass-eigenstate fields. We
begin by defining the gauge-eigenstate ``shifted" fields
\begin{equation}
H = \begin{pmatrix}
H^+ \cr
\frac{1}{\sqrt{2}}(v_H + H^0 + i \pi^0)
\end{pmatrix}~,
\hskip0.5in
T = \begin{pmatrix}
\frac{T^0 + v_T}{2} & -\,\frac{T^+}{\sqrt{2}} \cr
-\,\frac{T^-}{\sqrt{2}} & -\,\frac{(T^0 + v_T)}{2}
\end{pmatrix}~,
\label{eq:shifted}
\end{equation}
where $H^0$, $\pi^0$, and $T^0$ are real (neutral) fields, $H^\pm$ and $T^\pm$ are
complex charged fields, and we have chosen a convention for the sign of $T^\pm$
for later convenience.

Examining the field $\pi^0$, we see that no quadratic term arises from the potential
in Eq. \ref{eq:potentialii}, and therefore $\pi^0$ is the massless neutral Goldstone
boson ``eaten" by the $Z$-boson.
The only quadratic terms in the fields $H^\pm$ and $T^\pm$ arise from
the last term in  Eq. \ref{eq:potentialii}. Expanding, we find that only one linear
combination of $H^\pm$ and $T^\pm$ is massive
\begin{align}
h^\pm & = \frac{-2 v_T\, H^\pm + v_H\, T^\pm}{\sqrt{v^2_H + 4 v^2_T}} \nonumber \cr
& = -\sin\delta\, H^\pm + \cos \delta\, T^\pm~,
\end{align}
with
\begin{equation}
m^2_{h^\pm} = \frac{\mu}{4v_T}(v^2_H + 4 v^2_T) = \frac{\mu v_T v^2_H}{4 v^2_T}
+ \mu v_T~.
\label{eq:mpitilde}
\end{equation}
The orthogonal linear combination
\begin{equation}
\pi^\pm = \cos\delta\, H^\pm + \sin\delta\, T^\pm~,
\label{eq:pions}
\end{equation}
is massless, and corresponds to the Goldstone bosons ``eaten" by the
$W^\pm$.

The neutral scalar eigenstates are a bit more involved. The mass-squared matrix
in the $H^0$ -- $T^0$ basis is given by
\begin{equation}
{\mathsf m}^2   = \begin{pmatrix}
{1\over 2}\tilde{\lambda}_H v^2_H + 2 \mu v_T & 
\kappa v_H v_T-\frac{\mu v_T v_H}{2 v_T} \cr
\kappa v_H v_T-\frac{\mu v_T v_H}{2 v_T} &
\frac{\mu v_T v^2_H}{4 v^2_T} + \frac{1}{2} \lambda_T v^2_T
\end{pmatrix}~,
\end{equation}
where we have grouped factors of $\mu v_T$ to facilitate expanding in
powers of $v_T/v_H$.
Defining a mixing angle $\gamma$ \cite{Chen:2005jx,Chen:2006pb}, 
the lighter $(h_1)$ and heavier $(h_2)$ neutral scalar mass eigenstates are given by 
\begin{align}
h_1 & = \cos\gamma\, H^0 + \sin\gamma\, T^0 ~, \cr
h_2 & = -\,\sin\gamma\, H^0 + \cos\gamma\, T^0~,
\end{align}
with
\begin{align}
m^2_{h_1} & = \frac{1}{2} \tilde{\lambda}_H v^2_H + \mu v_T + \ldots~, 
\label{eq:mh1}\\
m^2_{h_2} & = \frac{\mu v_T v^2_H}{4 v^2_T} + \mu v_T + \frac{1}{2}
v^2_T (4 \tilde{\lambda}_H + \lambda_T - 8 \kappa+\frac{8\mu v_T}{v^2_H})+\ldots~,
\label{eq:mh2}
\end{align}
and
\begin{equation}
\sin\gamma = \frac{2v_T}{v_H} \left(1+
\frac{2 v^2_T}{\mu v_T}\left[\tilde{\lambda}_H - \kappa+\frac{\mu v_T}{v^2_H}\right] +\ldots\right)~.
\label{eq:singamma}
\end{equation}

Comparing Eqs. \ref{eq:sindelta} and \ref{eq:singamma} we see that the
mixing angles of the charged $(\delta)$ and neutral $(\gamma)$ states differ only starting at order $v^3_T/v^3_H$, while from Eqs. \ref{eq:mpitilde}
and \ref{eq:mh2} we see that the masses of the heavy charged and neutral states differ only starting at order $v^2_T$. Hence, to leading order in $v_T/v_H$, the linear combination of
doublet and triplet fields that becomes heavy is the same for both the charged and neutral scalars. 
The heavy fields are, to leading order, simply the Higgs triplet fields, which, according to Eq. \ref{eq:msquaredT}, have a mass of order $\mu v^2_H/4v_T$ in the small $v_T$ limit.
The reason for this behavior will become apparent in the following section (see the discussion following Eq. \ref{eq:relation}).

\subsection{Constructing the Low-Energy Effective Theory}
\label{sec:lowenergytheory}

Consider the equations of motion arising from the Lagrangian including the potential
in Eq. \ref{eq:potentialii}. The linear terms arising from the potential 
in the equations of motion for $H$ will be, at most, of order $v^2_H$. By contrast,
the linear terms arising from the potential in the equations of motion for $T$ will
receive contributions of order $\mu v^2_H/v_T$ from the last term in
Eq. \ref{eq:potentialii}. To leading order in $v_T/v_H$, therefore, the equations of motion 
reduce to the constraints
\begin{align}
\frac{\mu v_T v_H}{v^2_T}  \left[\frac{v_T}{v_H} H H^\dagger  + \frac{v_H}{2} T - \frac{v_H v_T}{4}\,{\cal I} \right] &  + \alpha\, {\cal I}   = 0 \\
{tr}\, T & = 0 
\end{align}
arising, respectively, from the $T$ equation of motion and the Lagrange multiplier. Solving these equations,
we find
\begin{equation}
\alpha = -\,\frac{\mu }{2}\left[H^\dagger H - \frac{v^2_H}{2}\right]~,
\end{equation}
and therefore
\begin{equation}
T = \left[ -\,\frac{2 v_T}{v^2_H}\, H H^\dagger + \frac{v_T}{v^2_H} \, (H^\dagger H)\,
{\cal I} \right] \left(1 + {\cal O}\left(\frac{v^2_T}{v^2_H}\right)\right)~.
\label{eq:relation}
\end{equation}

This information allows us to formally identify the states present in the low-energy theory.
Expressing Eq. \ref{eq:relation} in terms of the post-symmetry-breaking fields of Eq. \ref{eq:shifted}, we
find that the constant terms cancel, yielding
\begin{align}
T^0 - \frac{2 v_T}{v_H} H^0 & = \frac{2 v_T}{v^2_H} \left(-H^+ H^- + \frac{1}{2}({\pi^0})^2 + \frac{1}{2} ({H^0})^2\right)
+ \ldots \\
T^\pm - \frac{2 v_T}{v_H} H^\pm & = \frac{2 v_T}{v^2_H} \left(H^+ H^0 \pm i H^\pm \pi^0\right) + \ldots~.
\end{align}
Hence, the equations of motion (to this order in $v_T/v_H$) ensure that the linear
combinations of neutral and charged fields on the left hand side of these equations do not
produce single-particle states.  In other words, these combinations are the heavy states that are integrated out of the low-energy theory and $\sin\delta = \sin\gamma = 2 v_T/v_H$, in agreement with the discussion presented above in Sec. \ref{sec:massmix}.  It is the states orthogonal to these heavy states that are present in the low-energy effective theory.

Next, we can insert the leading-order solution to the equations of motion,  Eq. \ref{eq:relation}, 
 into the doublet-triplet Lagrangian to actually construct the effective 
low-energy theory that arises from integrating out the heavy states, up to corrections of
order $v^4_T/v^4_H$. The leading contribution to the low-energy potential
is
\begin{equation}
V(H)_{eff} = \frac{\left[\tilde{\lambda}_H + \frac{2 \mu v_T}{v^2_H}\right]}{4}
\left(H^\dagger H - \frac{v^2_H}{2}\right)^2 + \ldots~,
\label{eq:lambdahi}
\end{equation}
where the ellipses refer to terms of higher dimension, and higher order in $v^2_T/v^2_H$.
At this point, it is instructive to compare Eqs. \ref{eq:lambdah} and \ref{eq:lambdahi}. Note
that the four-point doublet coupling in the high-energy theory is given by $\lambda_H $, whereas the coupling
strength in the low-energy theory is $\lambda_H-2\mu v_T/v^2_H$. 
As anticipated, in order for the four-point couplings of the doublet
to be perturbative at both low- and high-energies, 
we must require that ${\lambda}_H$ and $\mu v_T/v^2_H$ should {\it each} be smaller
than $(4\pi)^2$ for both the low- and high-energy theories to remain
perturbative. In particular,
in the small-$v_T$ limit $\mu$ cannot grow faster than $1/v_T$.

The most interesting additional terms that arise from inserting Eq. \ref{eq:relation} into the doublet-triplet Lagrangian are those that
affect the $W$- and $Z$-boson masses:
\begin{equation}
\Delta{\cal L}_{WZ-masses} = \frac{4 v^2_T}{v^4_H} \left[
2 (H^\dagger H) D^\mu H^\dagger D_\mu H + (H^\dagger D^\mu H)(H^\dagger D_\mu H)
+ (\{D^\mu H^\dagger\} H)(\{D_\mu H^\dagger\} H)
\right]~.
\label{eq:deltalag}
\end{equation}
Combining these with the canonical $H$ kinetic energy term in Eq. \ref{eq:lagrangian} reproduces the
$W$- and $Z$-boson masses of Eq. \ref{eq:mwz}. We note that the second term of Eq. \ref{eq:deltalag} violates custodial symmetry \cite{Sikivie:1980hm,Weinstein:1973gj} and is responsible for the non-zero value of  $\Delta \rho$ \cite{Buchmuller:1985jz} in the low-energy effective theory. 

Finally, we note one subtlety in calculating in the low-energy theory: 
having integrated out $T$, the field $H$ in the 
effective theory constructed above
represents an appropriate ``interpolating field" in the low-energy theory (in the sense that
it has a non-zero amplitude to create all of the light one-particle scalar states), but it is
neither correctly normalized nor meant to be identified with the canonical $H$ field of the
high-energy theory described in Sec.\ref{sec:model} . In particular, 
below the scale of electroweak symmetry breaking (see Eq. \ref{eq:shifted})  the operator in Eq. \ref{eq:deltalag} in the low-energy theory includes a term of the form
\begin{equation}
\Delta{\cal L}_{WZ-masses} \supset \frac{4 v^2_T}{v^2_H} \partial^\mu H^\dagger \partial_\mu H~.
\end{equation}
Therefore, to this order in $v_T/v_H$, properly normalizing
the low-energy field $h_1$ requires being mindful of the relationship
\begin{equation}
h_1 = \left(1+\frac{2 v^2_T}{v^2_H} + \ldots\right) H^0~.
\label{eq:normalize}
\end{equation}

\subsection{Higher Loop-Order and Dimensional Analysis}
\label{sec:higher}

The calculations above have constructed the tree-level low-energy
effective Lagrangian. The effective Lagrangian at higher loop-order will
include terms of the same form: in order to construct the effective
theory to higher loop-order, we must ``match" the high-energy
and low-energy theories at the appropriate order in perturbation theory, choosing a renormalization
scale $Q$ of order $ m_{h^\pm,h_2}$ (as discussed, for example, in \cite{Manohar:1996cq}). Below the scale $Q =m_{h^\pm,h_2}$, the
parameters in the low-energy theory (including  $v_T$ as defined in terms of the coefficient of the custodial symmetry violating term in Eq. \ref{eq:deltalag}) 
only run due to the small, perturbative, dimension-four interactions in the low-energy 
theory --- namely, the gauge-couplings  and quartic Higgs-couplings. 
Because these corrections are small in a perturbative theory, 
the phenomenologically relevant issue for custodial
symmetry violation is the size 
of $v_T(Q=m_{h^\pm,h_2})=v^{ren}_T$,  {\it i.e.}, the size of the renormalized triplet vev as calculated
in the high-energy theory \cite{Chankowski:2006hs,Chen:2005jx,Chen:2006pb}. Similarly,
the value of $v_H$ relevant in the ${\cal O}(p^2)$ terms in the effective Lagrangian
(Eq. \ref{eq:lambdahi}) is the value $v_H(Q=m_{h^\pm,h_2})=v^{ren}_H$, and
the value of the triplet contribution to the rho parameter, $\Delta \rho_T
$, is given by $4 (v^{ren}_T)^2/(v^{ren}_H)^2$, the same 
expression as in Eq. \ref{eq:deltarho} with $v_{T,H} \to v^{ren}_{T,H}$. Phenomenologically,
therefore, we see that $v^{ren}_T \ll v^{ren}_H$ and, numerically,  $v^{ren}_H \approx v = 
(\sqrt{2} G_F)^{-1/2}$.

The low-energy effective theory is non-renormalizable and cannot be a fundamental
theory. In general, we expect a low-energy theory to be valid only below some scale $\Lambda$, 
where new physics associated with a high-energy completion \cite{Chivukula:1992gi} becomes relevant. In the case
of the doublet-triplet model, the high-energy completion corresponds to the exchange of the
heavy triplet-scalars,  $h_2$ and $h^\pm$, 
and we expect $\Lambda \simeq m_{h_2,h^\pm}$.
However, we note that the expansion parameter in the low-energy effective theory
is $(v^{ren}_T)^2/(v^{ren}_H)^2$ --- which is not obviously related to an expansion suppressed
by masses ($m^2_{h^\pm,h_2}$) of the heavy particles  
\cite{Chen:2005jx,Chen:2006pb,Chankowski:2006hs,Chankowski:2007mf}, as would be the normal expectation
\cite{Grinstein:1991cd}. 

Using  dimensional 
analysis,  we may estimate the upper bound on the
energy scale at which this low-energy theory breaks down. As shown
in \cite{Georgi:1985hf}, an effective theory of a scalar particle ($H$ in
this case) is determined by two dimensional constants:  the analog of
the pion-decay constant in the QCD chiral Lagrangian ($f$), and the  cutoff 
scale the low-energy theory ($\Lambda$).  The coefficient of the higher-dimensional term in Eq. \ref{eq:deltalag} should be of order $1/f^2$, and hence we find
\begin{equation}
f \simeq \frac{(v^{ren}_H)^2}{2 \sqrt{2}\, v^{ren}_T}~.
\end{equation}
Dimensional analysis in the low-energy theory imposes the constraint 
\cite{Weinberg:1978kz,Chivukula:2000mb}
$\Lambda \lesssim 4 \pi f$, with this inequality saturated only if the low-energy
theory is strongly-coupled. Using this inequality we find
\begin{equation}
\Lambda\simeq m_{h_2,h^\pm} \lesssim \frac{\sqrt{2}\, \pi (v^{ren}_H)^2}{v^{ren}_T} \approx \frac{2\sqrt{2} \pi v}{\sqrt{\Delta \rho_T
}}~.
\label{eq:bound}
\end{equation}
This expression provides a bound on $m_{h_2,h^\pm}$ which depends on the value
of $v^{ren}_T$ (or, equivalently, $\Delta \rho_T
$) in the low-energy theory,
a bound that will be crucial to our discussion
in Sec. \ref{sec:decoupling} of the fine-tuning required in the high-energy theory.
In the next section, we will establish a more precise bound on the masses $m_{h_2,h^\pm}$.

\section{Unitarity in Elastic $W_L^+W_L^-$ Scattering}

\label{sec:unitarity}

\begin{figure}
\begin{center}
\includegraphics[scale=0.45]{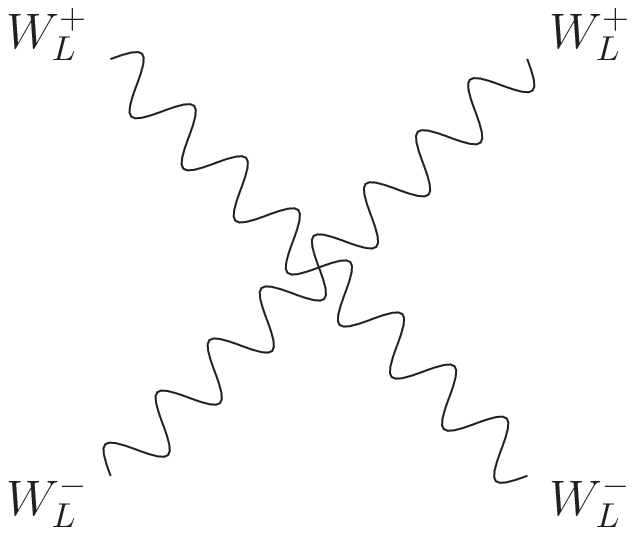}
\hskip+.75cm
\raise+1.1cm\hbox{+}
\hskip+.75cm
\includegraphics[scale=0.45]{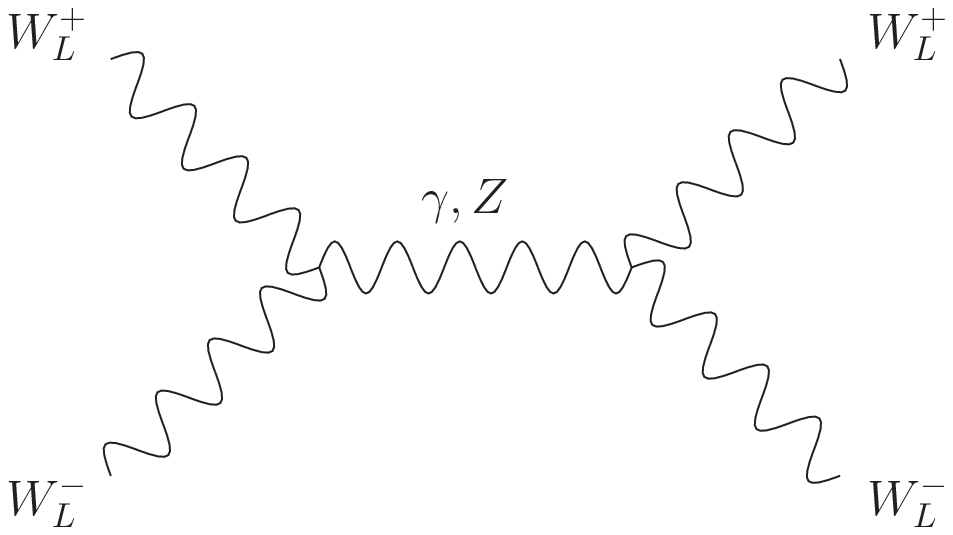}
\hskip+.75cm
\raise+1.1cm\hbox{+}
\hskip+.75cm
\lower+0.6cm\hbox{\includegraphics[scale=0.45]{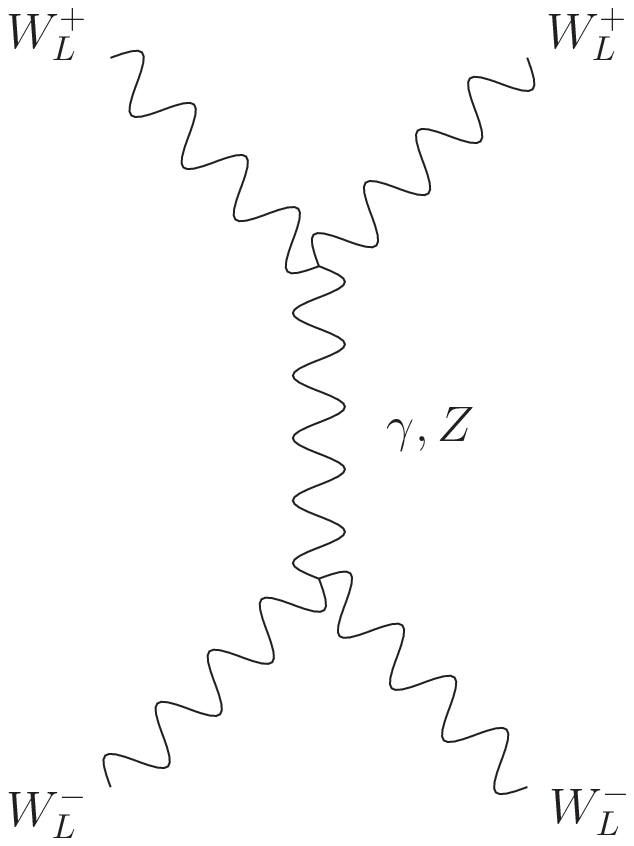}}
\end{center}
\caption{\label{fig:gauge-WW}Gauge interaction contributions to $W^+_LW^-_L$
scattering.}
\end{figure}

\begin{figure}
\begin{center}
\includegraphics[scale=0.45]{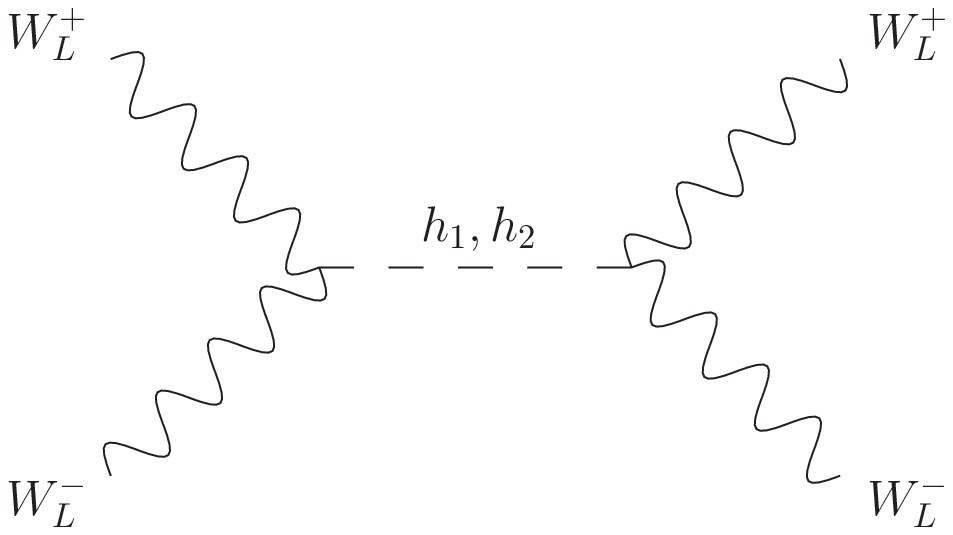}
\hskip+.75cm
\raise+1.1cm\hbox{+}
\hskip+.75cm
\lower+0.6cm\hbox{\includegraphics[scale=0.45]{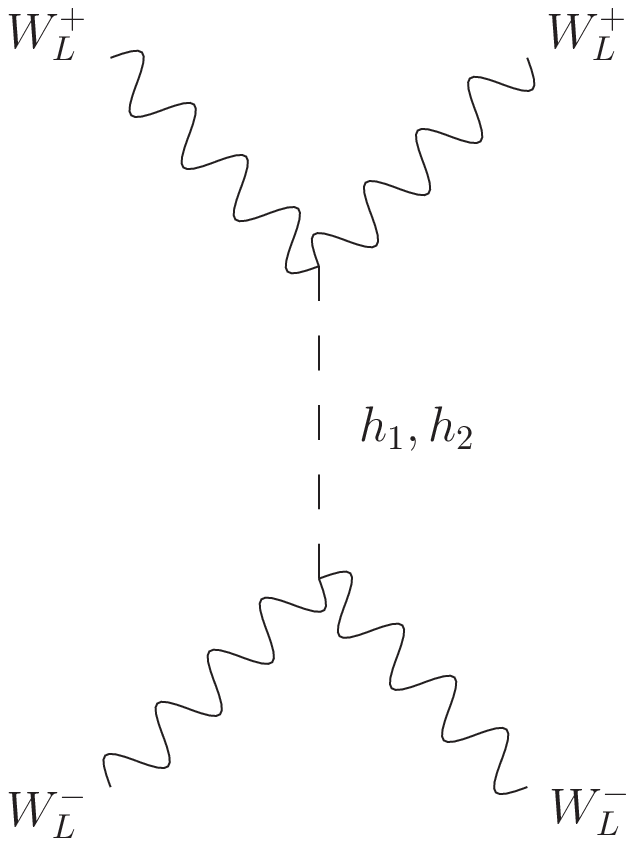}}
\end{center}
\caption{\label{fig:scalar-WW}Scalar exchange contributions to $W^+_LW^-_L$
scattering.}
\end{figure}

To delineate the connection between the masses of the heavy scalars ($h^\pm$
and $h_2$) and the size of $\Delta \rho_T
$, we turn to a calculation of elastic $W_L^+ W_L^-$
scattering.\footnote{This analysis is analogous to the unitarity bounds on the Higgs-boson
mass in the standard model \protect\cite{Dicus:1992vj,Veltman:1976rt,Lee:1977yc}.}  
We begin by considering $W^+_L W^-_L$ scattering at tree-level in the 
full doublet-triplet theory. Then, using the results of the previous section, we show
how these calculations are modified when working to higher-loop order.

\subsection{The Tree-Level $W^+_L W^-_L$ Scattering Amplitude}

At tree-level, $W^+_L W^-_L$  scattering arises both from 
gauge-boson self-interactions (Fig. \ref{fig:gauge-WW})
and from scalar exchange (Fig. \ref{fig:scalar-WW}). The gauge-boson self-interactions
are precisely the same as in the standard model, while the relevant gauge-scalar couplings
are
\begin{align}
g_{h_1WW} &= \frac{e^2v_H}{2s_W^2}+\frac{3e^2v_T^2}{s_W^2v_H}+\ldots
\label{eq:gh1ww}\\
g_{h_2WW} &= \frac{ e^2v_T}{s_W^2}+\frac{2e^2v_T^3\left((\kappa-\tilde{\lambda}_H)v_H^2-3\mu v_T\right)}{s_W^2v_H^2\mu v_T}+\ldots~,
\label{eq:gh2ww}
\end{align}
where, as before, $s_W$ is the sine of the weak mixing angle.
As in the standard model, the leading $E^4$ growth in the scattering amplitude arising from the separate gauge self-interaction diagrams of Fig. \ref{fig:gauge-WW} cancels when the diagrams are summed.  The most dangerous growth, therefore, occurs at order $E^2$. 

In the standard model, this order $E^2$ growth
in the four-point and gauge-boson-exchange contributions to the scattering amplitude is cancelled entirely by the effects of Higgs-boson exchange. 
However for the doublet-triplet model, Eq. \ref{eq:v} implies
\begin{equation}
v \approx v_H + \frac{2 v^2_T}{v_H}~,
\end{equation}
and, therefore,
\begin{equation}
g_{h_1WW} = g^{SM}_{hWW} + \frac{2 e^2 v^2_T}{s^2_W v_H} + \ldots~,
\label{eq:h1WW}
\end{equation}
where $g^{SM}_{hWW}$ is the standard model higgs-$WW$ coupling.
Due to the ${\cal O}(v^2_T/v_H^2)$ correction in Eq. \ref{eq:h1WW}, we expect that $h_1$ exchange
alone {\it will not} cancel the ${\cal O}(E^2)$ growth in the tree-level scattering amplitude of the doublet-triplet model  -- a property we will now demonstrate explicitly.

We consider first the high-energy tree-level amplitude, 
in the regime $E_{CM} \gg m_{h_1,h_2}$. The ${\cal O}(E^2)$ piece of the scattering amplitude
in this regime is
\begin{equation}
\label{M16}
\mathcal{M}(W_L^+ W_L^- \to W_L^+ W_L^-) \simeq \frac{E_{CM}^2(1+c)}{8M_W^4}\left[
\frac{e^2}{s_W^2}\left(4M_W^2-3M_Z^2c_W^2\right)
- g_{h_1WW}^2 - g_{h_2WW}^2 \right]~,
\end{equation}
where $c = \cos\theta_{CM}$,  and $E_{CM}$ and $\theta_{CM}$ are the center of mass
energy and scattering angle respectively.
It is easy to verify that the gauge-scalar couplings satisfy the sum rule
\begin{equation}
g_{h_1WW}^2 + g_{h_2WW}^2
= \frac{e^2}{s_W^2}\left(4M_W^2-3M_Z^2c_W^2\right)~,
\label{eq:sumrule}
\end{equation}
and, therefore, exchange of the two neutral scalars $h_{1,2}$ unitarizes $WW$
scattering at high-energies.

Next, consider the low-energy region $m_{h_1,h_2} \gg E_{CM} \gg M_{W,Z}$.
In this limit, none of the scalars contribute, and we only have the contribution of the gauge bosons; 
 the ${\cal O}(E_{CM}^2)$ amplitude is given by,
\begin{equation}
\mathcal{M}(W_L^+ W_L^- \to W_L^+ W_L^-) \simeq \frac{e^2E_{CM}^2(1+c)}{8M_W^2s_W^2}
\left(4-\frac{3}{\rho}\right)=-\,\frac{e^2\,u}{4M_W^2s_W^2} \left(4-\frac{3}{\rho}\right),
\end{equation}
where $u$ is the $u$-channel center of mass energy-squared. This expression
agrees with the general low-energy theorem for $W_L^+ W_L^- \to W_L^+ W_L^-$ 
in  \cite{Chanowitz:1986hu,Chanowitz:1987vj}.

Finally, consider the intermediate regime $m_{h_2} \gg E_{CM} \gg m_{h_1},\, M_{W,Z}$ --
the regime in which the low-energy theory of the previous section applies. In this
regime $h_2$-exchange does not contribute, and the cancellation implied by the sum-rule
of Eq. \ref{eq:sumrule} is incomplete. We find
\begin{equation}
\mathcal{M}(W_L^+ W_L^- \to W_L^+ W_L^-) \simeq \frac{E_{CM}^2(1+c)}{8M_W^4}
\left[
\frac{e^2}{s_W^2}\left(4M_W^2-3M_Z^2c_W^2\right)
- g_{h_1WW}^2 \right] = \frac{E_{CM}^2(1+c)}{8M_W^4}\,g^2_{h_2WW}~,
\label{eq:esquared}
\end{equation}
where we have used Eq. \ref{eq:sumrule} to simplify the result.
Due to the growth in this amplitude, there is an upper bound on $E_{CM}$ whose value depends on $g^2_{h_2 WW}$. We elaborate on this next.

\subsection{Tree-Level Unitarity and Bounds on $m_{h^\pm,h_2}$}

Using Eq. \ref{eq:esquared}, we find the spin-0 partial wave scattering
amplitude
\begin{equation}
a_0(W_L^+ W_L^- \to W_L^+ W_L^-) = \frac{1}{32\pi} \int^{+1}_{-1}
d\cos\theta_{CM} \,\mathcal{M}(W_L^+ W_L^- \to W_L^+ W_L^-)
 \simeq \frac{g_{h_2WW}^2s}{128\pi M_W^4}~,
\end{equation}
where $s=E^2_{CM}$. 
To satisfy partial wave unitarity, this tree-level amplitude
must be less than 1/2, the maximum value for the real
part of any amplitude lying in the Argand circle. 

From the sum rule in Eq. \ref{eq:sumrule}, we see that inclusion of $h_2$ exchange
is required for perturbative unitarity to be restored. Requiring that the low-energy
theory of Section \ref{sec:lowenergytheory} remain perturbative,
therefore, results in an upper bound on the mass of $m_{h_2}$
\begin{equation}
m_{h_2} \lesssim \frac{8\sqrt{\pi} M^2_W}{g_{h_2 WW}}~.
\label{eq:boundi}
\end{equation}
By appling Eq. \ref{eq:gh2ww} we obtain the tree-level bound
\begin{equation}
m_{h^\pm,h_2}  \lesssim \frac{2 \sqrt{\pi} v^2_H}{v_T}~.
\label{eq:unitarityboundtree}
\end{equation}

\subsection{Unitarity at Higher Loop-Order}

While these results have been derived at tree-level, our discussion of the effective low-energy
theory in the previous section allows us to generalize to higher-loop order. From Eq. \ref{eq:esquared},
we see that the relevant couplings in the low-energy Lagrangian are the gauge-couplings
and $g_{h_1WW}$. Taking into account the wavefunction normalization of Eq. \ref{eq:normalize},
we see that the coupling $g_{h_1 WW}$ is reproduced in the low-energy theory from a combination of
the kinetic energy terms in Eq. \ref{eq:lagrangian} and the custodial symmetry violating
terms in Eq. \ref{eq:deltalag}. From our discussion in Sec. \ref{sec:higher}, therefore, we see
that the effects of higher-loop order corrections in the high-energy theory can be summarized
by the replacements $v_{T,H} \to v^{ren}_{T,H}$. We conclude that the bound
in Eq. \ref{eq:unitarityboundtree} becomes
\begin{equation}
m_{h^\pm,h_2}  \lesssim \frac{2 \sqrt{\pi} (v^{ren}_H)^2}{v^{ren}_T} \approx
 \frac{4 \sqrt{\pi}\, v }{\sqrt{\Delta\rho_T
}}~,
\label{eq:unitaritybound}
\end{equation}
or
\begin{equation}
v^{ren}_T \lesssim \frac{2 \sqrt{\pi} v^2}{m^2_{h^\pm,h_2}}~,
\label{eq:unitarityboundii}
\end{equation}
in the limit $v^{ren}_T \ll v^{ren}_H\approx v$. 
This bound agrees parametrically with that anticipated in
Eq. \ref{eq:bound}.

An alternative interpretation for Eq. \ref{eq:unitaritybound} is obtained by using 
Eqs. \ref{eq:mpitilde} and \ref{eq:mh2} in the low-energy theory, from which we obtain the inequality
\begin{equation}
\frac{1}{(4\pi)^2}\,\frac{\mu v^{ren}_T}{(v^{ren}_H)^2}  \lesssim \frac{1}{\pi}~.
\label{eq:unitarityboundi}
\end{equation}
Here again we see that the combination $\mu v^{ren}_T/(v^{ren}_H)^2$ behaves like a dimensionless
coupling, and the bound of Eq. \ref{eq:unitaritybound} insures that this coupling
remains perturbative.

Finally, we remark that a similar unitarity analysis can be completed
for $W^+_L W^-_L  \to Z_L Z_L$. In this case, in addition to $h_{1,2}$ exchange in the
$s$-channel, one must also include $h^\pm$ exchange in the $t$- and $u$-channels.
In the limit $v_T \ll v_H$, however, one finds that the ${\cal O}(E^2_{CM})$ amplitude
vanishes up to ${\cal O}(v^2_T/v^2_H)$. That is,  $h_1$-exchange suffices (to this order) 
to eliminate the growth in the scattering amplitude, and the process $W^+_L W^-_L
\to Z_L Z_L$ does not provide a stronger bound than Eq. \ref{eq:unitaritybound}.

\section{Non-Decoupling and Fine-Tuning}
\label{sec:decoupling}

\begin{figure}
\includegraphics[scale=0.5]{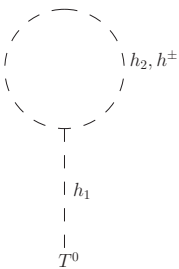}
\caption{\label{fig:tadpole} Dangerous tadpole diagrams which renormalize
the triplet vev. In this diagram, $T^0$ corresponds to the gauge-eigenstate
shifted neutral triplet field of Eq. \protect\ref{eq:shifted}, $h_1$ the tree-level light
neutral boson mass-eigenstate, and $h_2$ and $h_\pm$ the tree-level heavy
neutral and charged mass-eigenstates.}
\end{figure}

We are now ready to discuss the non-decoupling behavior of the triplet boson
demonstrated in \cite{Chen:2005jx,Chen:2006pb,Chankowski:2006hs}. The limit those references 
considered is  $v_T \to 0$ in the full (high-energy) theory.  From Eqs. 
\ref{eq:msquaredH} -- \ref{eq:lambdah} and \ref{eq:sindelta} we see that this amounts to the limit
$m^2_T \to \infty$ and $\sin\delta \to 0$ ($v_H \to v$), 
with $\mu v_T/v^2_H$, $\lambda_{H,T}$, and $\kappa$ remaining 
perturbative. At tree-level, $v_T$ in the full theory simply matches to $v_T$
in the low-energy theory --- and there are no residual effects at low-energy from
the heavy triplet bosons as $v_T \to 0$. At tree-level, therefore, the triplet decouples in this limit.

As discussed in Refs.  \cite{Chen:2005jx,Chen:2006pb,Chankowski:2006hs}, the
situation is different at one-loop. The issue is the contribution of the tadpole
diagrams illustrated in Fig. \ref{fig:tadpole}. In the $v_T \ll v_H$ limit, the trilinear
couplings in the diagrams with an internal $h_2$ or $h^\pm$ are equal to 
(at leading order)
\begin{equation}
g_{h_1 h^2_2} = g_{h_1 h^+ h^-}  = \kappa v_H + \frac{2 \mu v_T}{v_H} + \ldots ~,
\end{equation}
although these couplings differ at higher order. In addition to these couplings,
we will need the expression for $\sin\gamma$ in Eq. \ref{eq:singamma} and the
mass $m^2_{h_1}$ in Eq. \ref{eq:mh1}. These couplings and masses 
determine the relevant one-loop contribution to $\Delta v^2_T$:
\begin{equation}
\Delta v^2_T = \frac{ \mu v_T}{8\pi^2}\,
\frac{\left(\kappa v_H^2+2\mu v_T\right)}
 {\left(\tilde{\lambda}_Hv_H^2+2\mu v_T\right)}
\mbox{ln}\left(\frac{m_{h_2,h^\pm}^2}{Q^2}\right)+\ldots~,
\label{eq:nondecoupling}
\end{equation}
where $Q^2$ is the $\overline{MS}$ renormalization scale chosen in the computation.
The decoupling properties of the triplet at one-loop depend crucially on the behavior
of the ratio $\mu v_T / v^2_H$. On the one hand,  in the limit $v_T \to 0$ with $\mu$ fixed
(and therefore $\mu v_T \to 0$), this one-loop correction vanishes. In this case, the triplet
decouples at one-loop.\footnote{This result is consistent with
the decoupling theorem \protect\cite{Appelquist:1974tg}, as expected in the
case in which one takes only particle masses to be large.}
On the other hand, if one takes $\mu v_T/v^2_H$ fixed as $v_T \to 0$
\cite{Chen:2005jx,Chen:2006pb,Chankowski:2006hs}, then the one-loop correction
to $\Delta v^2_T$ does not automatically vanish and, in this sense, the triplet does not
decouple.

Note that the ``non-decoupling" limit depends crucially on the ratio of
{\it dimensionful} parameters $\mu v_T/v^2_H$ being held fixed. From 
Eqs.  \ref{eq:msquaredH} -- \ref{eq:lambdah}, we see that this non-decoupling
limit corresponds to taking {\it both} the parameters $m^2_T$ and $\mu$ large\footnote{
Note that both $m_T$ and $\mu$ are proportional to the GUT scale in a non-SUSY
$SU(5)$ theory \protect\cite{Chankowski:2007mf}. In particular, the dimensionful coupling
$\mu$ in the doublet-triplet theory can arise from a dimensionless coupling and a
symmetry-breaking scale.} (holding $m^2_T v^2_T/v^4_H$ and $\mu v_T/v^2_H$ fixed). Since a
dimensionful coupling ($\mu$) is becoming large in this limit, the absence of decoupling 
is {\it not} a violation of the decoupling theorem \cite{Appelquist:1974tg}.

The large tadpole contribution in the high-energy theory must, for a phenomenologically
acceptable theory with $\Delta \rho_T
 \ll 1$, be cancelled by an appropriately chosen counterterm
for $v_T$. Such a cancellation represents a fine-tuning in the high-energy theory.
Using Eq. \ref{eq:nondecoupling}  we see that the amount of fine-tuning is of order 
\begin{equation}
\frac{(v^{ren}_T)^2}{\Delta v^2_T} \equiv
\frac{v^2_T + \Delta v^2_T}{\Delta v^2_T} \simeq 
{\cal O}\left(\frac{8 \pi^2 v^{ren}_T}{\mu}\right)
\simeq {\cal O}\left(\frac{2\pi^2 v^2}{m^2_{h_2,h^\pm}}\right)~,
\label{eq:finetuning}
\end{equation}
where the last estimate derives from Eqs. \ref{eq:mpitilde} and \ref{eq:mh2} to
this order in perturbation theory.

Applying this result to a non-supersymmetric grand-unified theory \cite{Chankowski:2007mf},
where we expect\footnote{Making the Higgs-triplet light would require further fine-tuning,
but may be attractive in order to aid with coupling-constant unification and avoiding
proton decay constraints \protect\cite{Dorsner:2005fq}.} $m_{h_2,h^\pm} \simeq M_{GUT}$,
we see that the amount of fine-tuning required to keep $\Delta \rho_T
$ small 
is ${\cal O}(v^2_H/M^2_{GUT})$, 
precisely of the same form as the amount of fine-tuning that is required to maintain 
the weak scale/GUT scale
hierarchy \cite{Gildener:1976ai,Gildener:1976ih}. While this result may seem surprising,
it has a straightforward
interpretation in terms of the results of Sec. \ref{sec:unitarity}. From Eq. \ref{eq:unitaritybound},
we see that if the low-energy theory is to remain perturbative and if $m_{h_2,h^\pm} \approx M_{GUT}$, 
we must arrange for the properly renormalized low-energy triplet vev
 $v_T$ to be ${\cal O}(2\sqrt{\pi} v^2/M_{GUT})$. The fine-tuning in Eq. \ref{eq:finetuning}
 is a reflection of the fine-tuning required to lower the triplet vev from ${\cal O}(M_{GUT})$
 to this much lower size.  
 
Finally, we reiterate that as a consequence of the bound in Eq. \ref{eq:unitaritybound}, it is not 
sufficient for the low-energy triplet vev to be small enough to produce an
experimentally acceptable value of $\Delta \rho_T
$. Rather, 
in order for the low-energy theory to remain perturbative up to a scale of order $M_{GUT}$,
the properly renormalized value of $v_T$ must be ${\cal O}(v^2_H/M_{GUT})$ or
smaller.  This, in turn, constrains $\Delta\rho_T$ to be far smaller than the current experimental bound. Re-writing Eq. \ref{eq:unitaritybound} with $m_{h^\pm,h_2} \simeq M_{GUT}$, 
we find
\begin{equation}
\Delta \rho_T
 \lesssim \frac{16 \pi v^2}{M^2_{GUT}} \approx 3.1 \times 10^{-24}\,
\left(\frac{10^{15}\,{\rm GeV}}{M_{GUT}}\right)^2~.
\end{equation}
Hence, for a fine-tuned non-supersymmetric grand unified theory that is perturbative
at all energies, the presence of Higgs-triplet bosons with GUT-scale masses
is entirely irrelevant for low-energy electroweak phenomenology. 

In contrast \cite{Chen:2003fm}, for a little-Higgs model 
\cite{ArkaniHamed:2001nc,Schmaltz:2005ky} 
in which the scale of new physics is of order $M \approx 30$ TeV, the properly renormalized value of $v_T$ must be  ${\cal O}(v^2_H/M)$ or smaller.   Then 
assuming $m_{h^\pm,h_2} \simeq M$ one has 
\begin{equation}
\Delta \rho_T \lesssim \frac{16 \pi v^2}{M^2} \approx .003 \left(\frac{3 \times 10^4\,{\rm GeV}}{M}\right)^2~.
\end{equation}
In this case, the constraints on $\Delta\rho_T$ from perturbativity and experiment are 
comparable\footnote{Note that one must be careful in extracting electroweak limits in the
presence of new, custodially-violating, physics \protect\cite{Chen:2003fm,Chivukula:2000px,Peskin:2001rw}.} 
because the scale of new physics is much closer to the weak scale.

\section{Conclusions}

In this paper we have considered  the properties of a model incorporating
both a scalar electroweak Higgs doublet and an electroweak Higgs triplet. 
We constructed the low-energy effective theory below the scale of the triplet-mass, 
showing explicitly that the higher-dimensional, non-renormalizable, operators responsible
for deviations in $\rho$ are suppressed not by inverse powers of the triplet mass, but rather
by powers of the renormalized triplet vev ($v^{ren}_T$) divided by the 
renormalized doublet vev ($v^{ren}_H$).
We have demonstrated that perturbative unitarity in the low-energy theory
breaks down at a scale inversely proportional to the renormalized triplet vev,
both by using dimensional analysis and by an explicit computation of
$WW$ scattering in the low-energy theory.
We have shown that two possibilities remain: either the contribution of the triplet vev is comparable to the existing experimental bounds, in which case the triplet scalars must have a mass of order 30 TeV or
lower, or if the triplet masses are much larger than 30 TeV, as in the case of a non-supersymmetric
GUT theory,  then the triplet vev is too small to be phenomenologically relevant. 
We have also clarified that, despite the non-decoupling behavior of the Higgs-triplet, 
this model does not violate the decoupling theorem since it
incorporates a large dimensionful coupling.

Finally, we note an interesting parallel between this work and non-decoupling effects
in seesaw-extended MSSM models \cite{Dedes:2007ef}. 
The non-decoupling of the triplet in the doublet-triplet
Higgs model can be viewed as arising from the fact that the Goldstone bosons eaten
by the $W^\pm$ are combinations of doublet- and triplet-states, as shown in 
Eq. \ref{eq:pions}. Similarly, in a seesaw-extended MSSM it is possible to consider
a limit in which the low-energy sneutrino field remains partially the superpartner of
a sterile Majorana seesaw neutrino field -- even in the limit of large seesaw mass
\cite{haber}.

\section{Acknowledgements}
This work was supported in part by the US National Science Foundation under
grant  PHY-0354226. The authors acknowledge the hospitality and support of the
    Radcliffe Institute for Advanced Study, the CERN Theory group, and  the Galileo Galilei 
    Institute for Theoretical Physics during the completion of this work.
We thank Howard Haber, Ken Hsieh, J{\"u}rgen Reuter, and Carl Schmidt for useful discussions.

\end{document}